      [1999/12/01 v1.4c Il Nuovo Cimento]
\documentclass{cimento}
\usepackage[T1]{fontenc}
\usepackage[ansinew]{inputenc}

\begin{document}

\title{Formulation of an Evolutionary Quantum Cosmology}
\author{G.~Montani\from{ins:x}\from{ins:y}.}
\instlist{\inst{ins:x} ENEA C.R. Frascati (Dipartimento F.P.N.), via Enrico Fermi 45, 00044 Frascati, Rome, Italy.
\inst{ins:y} ICRA-International Center for Relativistic Astrophysics\\
Dipartimento di Fisica (G9), Universit\`a  di Roma, ``La Sapienza", Piazzale Aldo Moro 5, 00185 Rome, Italy.
 }
\PACSes{{83.C}}

\maketitle

\begin{abstract} 
We provide an evolutionary formulation of a generic
quantum cosmology. Our starting point is the request
that all quantities living on the slicing have to be 3-tensors.
This statement, when applied to the lapse function and the shift
vector, yields the no longer vanishing behavior of the
super-Hamiltonian. Then, we provide the formulation of
an evolutionary quantum cosmology in correspondence to a
generic Universe asymptotically to the initial singularity.

\end{abstract}

\section{General statements}

In recent works (see \cite{M02} and \cite{MM04a,MM04b}) was
presented a revised approach to the canonical quantum gravity,
which leads, via different procedures, to a Sch\"rodinger dynamics
for the 3-metric field. The cosmological implementation of this
evolutionary quantum gravity was then presented in
\cite{M03,BM06}.
The starting point of the reformulation is
that the (3+1)-slicing acquires a precise physical
meaning on a quantum level too, due to the light
cone preservation on a space-time foam. The price to be payed for this
issue is, essentially, in all the proposed schemes, the fixing of
a reference frame. On one hand, we have to deal with non-test fluids
which provide a physical slicing of the space-time and modify the
constraints of the theory. On the other hand, as far as the reference
frame is fixed before the quantization is performed, then the dynamics
acquires an evolutionary character. The result
of this new point of
view is determining a matter-time dualism for the quantum geometrodynamics.
Similar conclusions were inferred in \cite{BK95}, but there
they stressed mainly the first side of this dualism, i.e.
matter$\rightarrow$ time, while our approaches take more attention
to the opposite direction time $\rightarrow$ matter.\\
Here, we discuss, in some detail, the formulation of a
evolutionary quantum cosmology, as referred to a generic
inhomogeneous Universe.
The adopted procedure to reach a Schr\"odinger
dynamics relies on the request that all the quantities,
coming out from the slicing, behave as 3-tensors under the admissible
coordinates transformations. To ensure that the lapse function
and the shift vector obey the proper transformations, we have to restrict
ourselves to generic 3-diffeomorphisms, but to constant time
displacement only. The evolutionary implications
and the cosmological issues of such a restriction are then investigated.

In Section 2, we revise the Arnowitt-Deser-Misner
(ADM) \cite{ADM62}
formulation of General Relativity as viewed
in the vier-bein picture. Section 3 is devoted to
outline the dynamical implications of dealing with
a physical slicing. The request that all quantities
are 3-tensors is shown to remove the vanishing behavior
of the super-Hamiltonian. Section 4 provides an appropriate
formulation of the classical dynamics.
for a generic inhomogeneous Universe.
This scheme is then quantized in section 5
within the new evolutionary paradigm. 

\section{ADM-Formulation of the dynamics}

In the ADM representation of the geometrodynamics,
the space-time is sliced into a 1-parameter family of
spacelike hypersurfaces $\Sigma ^3_t$, defined by 
$t^{\mu } = t^{\mu }(t,\; x^i)$
($\mu = 0,1,2,3$ and $i = 1,2,3$). 
The deformation vector $N^{\mu }$ admits the decomposition
\begin{equation}
N^{\mu } \equiv \partial _tt^{\mu } =
Nn^{\mu } + N^i\partial _it^{\mu }
\label{a}
\, ,
\end{equation}
where $N$ and $N^i$ denote the lapse function and the
shift vector respectively, while the normal $n^{\mu }$
and the tangent fields $\partial _it^{\mu }$ fix
together a local basis of the manifold
$\mathcal{V}^4 = \Sigma ^3_t\times R$. A (3+1)-representation of the 4-metric field
$g_{\mu \nu }(y^{\rho })$
is reached by regarding
$t^{\mu } = t^{\mu }(t,\; x^i)$
as a coordinates transformation toward the new basis 
$\{ N^{\mu },\; \partial _it^{\mu }\}$, in which the
line element rewrites
\begin{equation}
ds^2 = g_{\mu \nu }dt^{\mu }dt^{\nu } =
-N^2dt^2 + h_{ij}(dx^i + N^idt)(dx^j + N^jdt)
\label{b}
\, ,
\end{equation}
Above,
$h_{ij}\equiv g_{\mu \nu }\partial _it^{\mu }
\partial _jt^{\nu }$ is the 3-metric induced on the
spatial hypersurfaces.\\
Taking a 3-bein representation of the 3-metric
$h_{ij} = \delta _{ab}u^a_iu^b_j$
($\delta _{ab}$
being the Euclidean metric referred to the bein
indexes), then 
the ADM-action of the gravitational field acquires the form
\begin{equation}
S_{ADM} = \int_{\Sigma^3_t\times R}dtd^3x
\left\{ \pi _a^i\partial _tu^a_i - NH - N^iH_i\right\} 
\label{e}
\, ,
\end{equation}
where, $\pi _a^i$ are the conjugate momenta to $u^a_i$.
The super-Hamiltonian $H$ and the supermomentum $H_i$ read 
respectively
\begin{eqnarray}
\label{f}
H = \frac{c^2\chi }{4u}\left\{2 \pi _a^i\pi _b^j
u^a_ju^b_i - \left( \pi _a^iu^a_i\right) ^2\right\} -
\frac{u}{2\chi }F^{ab}_{ij}u_a^iu_b^j\\
H_i = -u^a_iH_a =
-u^a_i\left(\partial _j\pi _a^j + {\omega ^b_{aj}}\pi_b^j\right)
\, ,
\end{eqnarray}
where $u\equiv det u^a_i$ and
${\omega ^{ab}}_i = (~^3\nabla _iu^a_k)u^k_b$
($~^3\nabla $ denoting the 3-covariant derivative)
is a 1-form connection.
$F^{ab}_{ij}$ is the 2-form curvature constructed
by such 1-form.

In the action (\ref{e}), $N$ and $N^i$ enter as Lagrangian
multipliers, whose variation leads to the constraints
$H = H_i = 0$.
We observe that
this vanishing nature of $H$ and $H_i$
respectively reflects the
time and spatial diffeomorphisms
invariance of the theory respectively. 
Finally, three additional constraints are requested
to kill the three redundant degrees of freedom
which the 3-bein has with respect to the 3-metric.
Such constraints take the form
$\pi ^i_au_{ib} - \pi ^i_bu_{ia} = 0$
and they correspond to the local $SO(3)$
invariance.
Below, we will account for these constraints in
the dynamics by a suitable choice pf the 3-bein
structure (see Section 4).

\section{Physics of the slicing dynamics}

Within the framework of
a physical separation between space and time,
the only admissible geometrical quantities
have to appear
in the form of 3-tensors (scalars, vectors, etc.).
Otherwise, the spirit
of General Relativity is violated
on the spatial hypersurfaces $\Sigma ^3_t$.
Since, under the time
diffeomorphism $t^{\prime } = f(t)$,
the lapse function $N$
and the shift vector $N^i$ are multiplied by the factor
$\partial _tf$, then we see that their 3-tensor nature is
restored as soon as we restrict ourselves to the (global)
time displacements $t^{\prime } = t + C$, $C$ being
a constant term. 
Instead, the generic space transformations 
${x^{\prime }}^i = {x^{\prime }}^i(x^l)$ naturally fulfill the
(3+1)-physics. Thus, in what follows, we analyze the
implications of the gravitational Lagrangian invariance under the
infinitesimal displacements
\begin{equation}
t^{\prime } = t + \xi
,\, \quad
x^{\prime ^i} = x^i + \xi ^i (x^l)
\, ,
\label{h}
\end{equation}
where $\xi$ is a constant quantity,
while $\xi ^i$ are generic ones.

Then, from the invariance of the dynamics under this class
of transformations and 
by few standard steps (in the spirit of the N\"other theorem),
we arrive to the key conservation law
(we stress that
$\pi_a^i{\omega ^a}_{bc}u^b_i = 0$ by virtue of the
constraints at the end of Section 2) 
\begin{equation}
\label{LCF}
\int_{\Sigma^3_t}d^3x
\left\{
\partial _t\left[ \pi _a^i\partial _tu^a_i - \mathcal{L}\right] 
- \left[ \partial _i\pi _a^i - {\omega ^b}_{ac}u^c_j\pi ^j_b\right]
\xi ^a\right\} = 
\int_{\Sigma^3_t}d^3x
\left\{ \partial _t\mathcal{H}                                                      
+ H_a\xi ^a\right\} = 0
\, ,
\end{equation}
$\mathcal{H}$ being the full Hamiltonian density.
Since the 3-dimensional General Relativity Principle
has to be preserved under the slicing, we have to require
that $H_a\equiv 0$ and the above constraint
simply rewrites
$\partial _tH = 0\; \Rightarrow \; H = E(x^l)$,
$E$ being a generic integration function, fixed by the
initial conditions on the gravitational system.\\
Thus, we see that the request to deal with 3-tensors only,
implies that the super-Hamiltonian constraint is removed.
Therefore, the action of the gravitational field
takes, for a physical (3+1)-observer, the final form
\begin{eqnarray}
\label{acv}
S_{ph} =
\int_{\Sigma^3_t}dtd^3x
\left\{
p_N\partial _tN +
\Xi ^ip_{N^i} +            
\pi _a^i\partial _tu^a_i - NH - N^iH_i\right]
\, , 
\label{CL}
\end{eqnarray}
where $\Xi^i$ are Lagrangian multipliers introduced
to emphasize the vanishing nature of the momenta
$p_{N^i}$ (being $p_{N}$ and
$p_{N^i}$ conjugate to $N$ and $N^i$ respectively).

\section{Dynamics of a generic Universe}

We now perform a significant reduction of the
variational
principle, as applied to a generic cosmology.
Below, we deal with an arbitrary reference frame and 
we analyze the dynamics 
asymptotically to the cosmological singularity.
To this end we observe that
a generic cosmological solution is represented
by the gravitational field having full degrees of
spatial inhomogeneity and whose
3-bein vectors are fixed in the form
\begin{equation}
u^a_i = \left( e^{Q}\right)^a_cO^c_b
\partial _iy^b    
\, , 
\label{bein}
\end{equation}
where, by the diagonal matrix
$Q_a^b = \frac{1}{2}q^a(t, x^l)\delta _a^b$
and $y^a(t, x^l)$, we determine six 3-scalars,
which resume the gravitational field dynamics.
$O^a_b=O^a_b(x^i)$ is a $SO(3)$ matrix and
it provides three spatial functions
available to the Cauchy problem. This choice for the
3-bein structure removes the three redundant components
in it contained.
The action for the gravitational field
is strongly simplified because, as shown in
\cite{BM04}, for any form of $N$ and $N^i$,
the super-momentum constraint can be solved by adopting
the variables $y^a$ as new spatial coordinates.\\ 
Hence, 
we rewrite the dynamics by means of the 
Misner variables %\{ \alpha ,\; \beta _{\pm}\}$
\cite{M69}, defined by the linear transformation 
\begin{equation}
q^1 = \alpha + \beta _+ + \sqrt{3}\beta _-
, \quad
q^2 = \alpha + \beta _+ - \sqrt{3}\beta _-
, \quad
q^3 = \alpha - 2\beta _+ 
 .
\label{misner}
\end{equation}
These new variables allow,
asymptotically to the singularity
($\alpha \rightarrow -\infty$), to rewrite the action
(\ref{CL}) in the form
\begin{eqnarray}
\label{finact}
S_{M} = \int_{\Sigma ^3_t\times\Re}dt d^3 y\left\{
p_N\partial _tN +
p_{\alpha }\partial_t \alpha
+ p_+\partial_t \beta _+
+ p_-\partial_t \beta _- 
- NH\right\}\\ 
H = \frac{c^2ke^{-3\alpha }}{3}\left[ 
-p_{\alpha }^2 + p_+^2 + p_-^2 \right]
- U(\alpha ,\; \beta _{\pm })\\
U = \frac{1}{2k\mid J\mid ^2}e^{\alpha } V(\beta _{\pm }), \quad
V(\beta _{\pm}) =
\lambda _1^2e^{4\beta _+ + 4\sqrt{3}\beta _-} +
\lambda _2^2e^{4\beta _+ - 4\sqrt{3}\beta _-} +
\lambda _3^2e^{-8\beta _+} 
\, .
\end{eqnarray}
Above, $J$ denotes the Jacobian of the transformation
from $x^i$ to $y^a$, while the spatial functions
$\lambda _i(y^a)$
($i=1,2,3$) fix the model inhomogeneity.\\
The most relevant feature of the obtained
dynamics
consists of the parametric role played here
by the spatial
coordinates.
In fact, it comes out because the potential terms
containing the gradients of
$\alpha$ and $\beta _{\pm}$
result to be asymptotically negligible.
Thus, the Wheeler Superspace of this
theory decouples into
$\infty ^3$ independent 3-dimensional minisuperspaces.

\section{Canonical quantization}

Since the total Hamiltonian of the system reduces,
near the singularity,
to the sum of $\infty ^3$ independent point-like
contributions, then the Schr\"odinger
functional equation splits correspondingly.
Thus, 
fixing the space point $y^a$,
the quantum dynamics reads
(we denote with the subscript $y$ any minisuperspace quantity)
\begin{eqnarray}
\label{sch}
i\hbar \partial _t \psi _y = 
\hat{H}_y\psi _y = \left\{ \frac{c^2\hbar ^2k}{3} \left[ 
\partial _{\alpha }e^{-3\alpha }\partial _{\alpha }
-e^{-3\alpha }\left( \partial ^2_+ + \partial ^2_-  
\right) \right] 
- \frac{3\hbar ^2}{8\pi } e^{-3\alpha }\partial ^2_{\phi }
\right\} \psi _y - \\ 
-\left( \frac{1}{2k\mid J\mid ^2}e^{\alpha } V(\beta _{\pm })
- \frac{\Lambda }{k}e^{3\alpha }\right) \psi _y,\qquad
\psi _y = \psi _y (t , \; \alpha , \; \beta _{\pm} , \; \phi )
\, .
\end{eqnarray}
Above, to make account of the inflationary scenario,
we included in the dynamics a massless
scalar field $\phi$ and a cosmological constant $\Lambda $.
The presence of these two terms allow to model the main features
of the inflaton field dynamics in the pre-inflation and
slow-rolling phases. 
We now take the following integral representation for the
wavefunction $\psi _y$
\begin{eqnarray*}
\label{exp}
\psi _y = \int 
d\mathcal{E}_y \mathcal{B}(\mathcal{E}_y)
\sigma _y(\alpha , \; \beta _{\pm } , \; \phi ,\; \mathcal{E}_y)
exp \left\{ -\frac{i}{\hbar }\int _{t_0}^tN_y\mathcal{E}_ydt^{\prime }
\right\},\qquad
\sigma _y = \xi _y(\alpha , \; \mathcal{E}_y)
\pi _y(\alpha , \; \beta _{\pm } , \; \phi )
\, , 
\end{eqnarray*}
where $\mathcal{B}$ is 
fixed by the initial condition at $t_0$.
Hence, we get the following reduced problems
\begin{eqnarray}
\label{eigenvp}
\hat{H}\sigma _y = \mathcal{E}_y \sigma _y\\ 
\left( -\partial ^2_+ - \partial ^2_-
- \frac{9}{8\pi c^2k } \partial ^2_{\phi }
\right) \pi _y
- \frac{3}{2c^2\hbar ^2k^2\mid J\mid ^2}e^{4\alpha } V(\beta _{\pm })\pi _y =
v^2(\alpha ) \pi _y\\ 
\left[ \frac{c^2\hbar ^2k}{3}\left( 
\partial _{\alpha }e^{-3\alpha }\partial _{\alpha }
+ e^{-3\alpha }v^2(\alpha ) \right) + 
\frac{\Lambda }{k}e^{3\alpha }\right] \xi _y
= \mathcal{E}_y\xi _y
\, ; 
\end{eqnarray}
Here, in deriving the equation for $\xi _y$, we neglected the
dependence of $\pi _y$ on $\alpha $ because, asymptotically to
the singularity ($\alpha \rightarrow -\infty$)
it has to  be of higher
order (i.e. we address a well grounded adiabatic
approximation).\\ 
Such an approximate description of the
early Universe quantum dynamics  
has a deep physical meaning,
corresponding to require that the
scale of the inhomogeneities is 
super-horizon sized and local homogeneity overlaps 
the notion of causality.
This fact allows us to deal
with an evolutionary quantum cosmology which preserves the
causal picture from a physical point of view.\\

We would like to thank Francesco Cianfrani, for
his valuable comments on different steps of the
paradigm here presented.

\end{document}